\documentclass[lettersize,journal]{IEEEtran}
\usepackage{amsmath,amsfonts,amssymb} 
\usepackage{algorithmic,algorithm} 
\usepackage{array,tabularx} 
\usepackage{graphicx} 
\usepackage{cite} 
\usepackage{url} 
\usepackage{verbatim} 
\usepackage{multirow} 
\usepackage{booktabs} 
\usepackage{ragged2e} 
\usepackage{adjustbox} 
\usepackage{textcomp} 
\usepackage{stfloats} 
\usepackage{xcolor} 
\usepackage{newfloat,listings} 

\usepackage{times} 
\usepackage{helvet} 
\usepackage{courier} 

\usepackage[caption=false,font=normalsize,labelfont=sf,textfont=sf]{subfig}

\begin{document}

\title{Senti-iFusion: An Integrity-centered Hierarchical Fusion Framework for Multimodal Sentiment Analysis under Uncertain Modality Missingness}

\author{Liling Li, Guoyang Xu, Xiongri Shen, Zhifei Xu, Yanbo Zhang, Zhiguo Zhang, Zhenxi Song
\thanks{
This work was supported in part by the National Natural Science Foundation of China under Grant Nos. 82272114 and 62306089, and in part by the Shenzhen Science and Technology Program under Grant Nos. RCBS20231211090800003 and ZDSYS20230626091203008.
\textit{(Corresponding author: Zhenxi Song.)}
}
\thanks{
Liling Li, Guoyang Xu, Xiongri Shen and Yanbo Zhang are with the School of Computer Science and Technology, College of Information Science and Technology, Harbin Institute of Technology, Shenzhen, Guangdong, China(email: 24S151033@stu.hit.edu.cn, 23S151099@stu.hit.edu.cn, xiongrishen@stu.hit.edu.cn, 25S151085@stu.hit.edu.cn).}
\thanks{
Zhifei Xu, Zhiguo Zhang and Zhenxi Song are with the 
School of Intelligence Science and Engineering, College of Artificial Intelligence, Harbin Institute of Technology, Shenzhen, Guangdong, China (email: 25B965022@stu.hit.edu.cn, zhiguozhang@hit.edu.cn,  songzhenxi@hit.edu.cn).}
}

\markboth{Journal of \LaTeX\ Class Files,~Vol.~14, No.~8, August~2021}%
{Shell \MakeLowercase{\textit{et al.}}: A Sample Article Using IEEEtran.cls for IEEE Journals}


\maketitle

\begin{abstract}
Multimodal Sentiment Analysis (MSA) is critical for human-computer interaction but faces challenges when the modalities are incomplete or missing. Existing methods often assume pre-defined missing modalities or fixed missing rates, limiting their real-world applicability. To address this challenge, we propose \textit{Senti-iFusion}, an integrity-centered hierarchical fusion framework capable of handling both inter- and intra-modality missingness simultaneously. It comprises three hierarchical components: Integrity Estimation, Integrity-weighted Completion, and Integrity-guided Fusion.
First, the \textit{\textbf{Integrity Estimation}} module predicts the completeness of each modality and mitigates the noise caused by incomplete data.
Second, the \textit{\textbf{Integrity-weighted Cross-modal Completion}} module employs a novel weighting mechanism to disentangle consistent semantic structures from modality-specific representations, enabling the precise recovery of sentiment-related features across language, acoustic, and visual modalities. 
To ensure consistency in reconstruction, a dual-depth validation with semantic- and feature-level losses ensures consistent reconstruction at both fine-grained (low-level) and semantic (high-level) scales.
Finally, the \textit{\textbf{Integrity-guided Adaptive Fusion}} mechanism dynamically selects the dominant modality for attention-based fusion, ensuring that the most reliable modality, based on completeness and quality, contributes more significantly to the final prediction.
\textbf{Senti-iFusion} employs a progressive training approach to ensure stable convergence. Experimental results on popular MSA datasets demonstrate that \textbf{Senti-iFusion} outperforms existing methods, particularly in fine-grained sentiment analysis tasks.
The code and our proposed Senti-iFusion model will be publicly available.
\end{abstract}


\section{Introduction}

\begin{figure}[t]
\centering
\includegraphics[width=1.0\columnwidth]{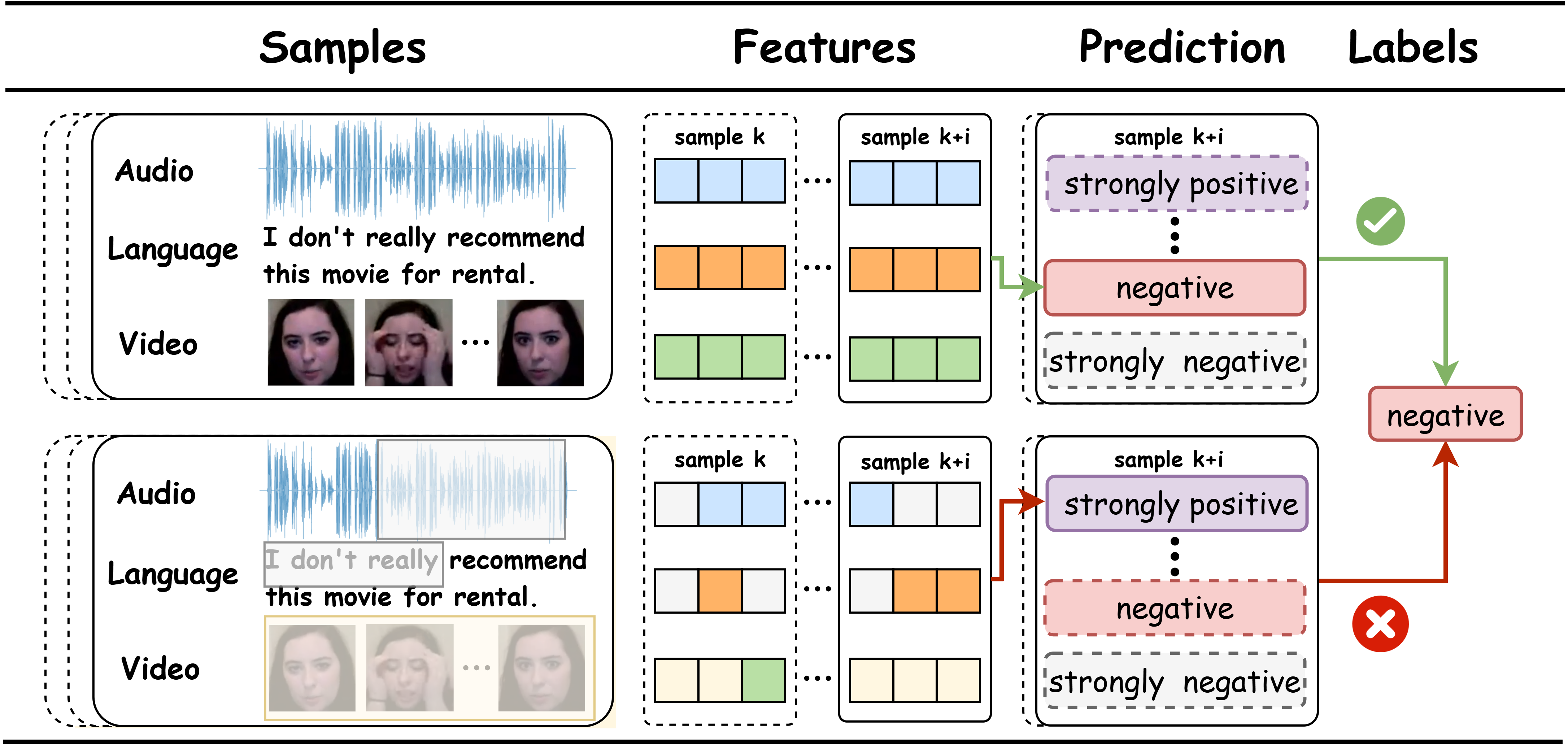} 
\caption{
Illustration of a wrong prediction under a dual-level missingness scenario, where missing data across multiple modalities leads to incorrect sentiment classification.
}
\label{fig: intro}
\end{figure}

 \IEEEPARstart{D}espite recent progress in multimodal sentiment analysis (MSA), practical deployment remains challenging due to uncertain modality missingness caused by real-world noise and sensor unreliability.
 Specifically, entire modalities may be absent (\emph{inter-modality missingness}), while partial segments within modalities may also be corrupted (\emph{intra-modality missingness}). These two forms of missingness often occur simultaneously, presenting a critical challenge for robust MSA. As illustrated in Figure~\ref{fig: intro}, such dual-level missingness can easily mislead standard models, making it essential to develop methods that handle both intra- and inter-modality missingness.

Current MSA approaches primarily handle missing data through two main strategies: (i) reconstructing absent modalities \cite{hrlf2024, yao2024drfuse, smil2021, du2018semi}, and (ii) modeling modality-specific deficiencies via adversarial learning \cite{yuan2023NIAT, cai2018deep, ganin2015unsupervised, goodfellow2014generative}.  
Joint learning frameworks \cite{li2024UMDF, LNLN2024} adopt unified architectures to address both inter- and intra-modality missingness simultaneously.  
Meanwhile, prompt-based approaches \cite{lang2025retrievalaugmented, mplmm2024, li2021prefix} exploit lightweight tuning to adapt models to specific missing modalities without modifying the backbone.

However, these approaches have notable limitations. LNLN \cite{LNLN2024} typically addresses missing data by integrating multimodal information but often focuses solely on within-modality missingness, neglecting the simultaneous occurrence of inter-modality missingness. Traditional cross-modal generation methods like HRLF\cite{hrlf2024} and DrFuse \cite{yao2024drfuse}, while capable of feature extraction and reconstruction, fail to fully leverage known missing data to guide the completion process. Additionally, the inherent heterogeneity between modalities introduces noise, degrading completion quality. The prompt tuning approach MPLMM \cite{mplmm2024}, though effective for modality-specific missingness, does not account for the coexistence of both within-modality and inter-modality missingness and is highly dependent on the quality of pre-trained models and datasets. As a result, existing methods fail to comprehensively address the complex multimodal missing issue and require a solution that handles both inter- and intra-modality missingness.

To address the challenges outlined above, we propose an Integrity-centered Hierarchical Fusion Framework called \textbf{Senti-iFusion} for MSA under uncertain data-missing scenarios. Our approach introduces stochastic input corruption to simultaneously simulate both intra-modality and inter-modality missingness. Each modality's complete features are randomly masked by the missingness procedure and then passed through an \textit{Integrity Estimation} module, which computes their respective integrity scores. The incomplete embeddings are then processed by an \textit{Integrity-weighted Cross-modal Completion} module. These embeddings are first disentangled into shared sentiment semantic features and modality-specific features, which are then corrected based on the estimated integrity scores, dynamically adjusting the composition of the multimodal features. The cross-modally completed surrogate representations are subsequently validated through a dual-depth scheme, enhancing both the accuracy and robustness of the missing feature recovery process.

To validate the effectiveness of the cross-modal completion representations, we propose a dual-depth validation scheme that evaluates completion quality at both the structural consistency and feature similarity levels.This mechanism computes Mutual Information (MI) loss and Mean Squared Error (MSE) loss for the decoded shared sentiment semantics and the overall features separately. The completion process minimizes the discrepancy between the original features and the reconstructed ones, both in terms of feature distribution and numerical values.

Unlike traditional methods that rely on a fixed dominant modality\cite{LNLN2024}, our approach introduces an \textit{Integrity-guided Adaptive Fusion} module, which dynamically selects the dominant and auxiliary modalities based on adaptively estimated integrity scores. This enables the model to choose the optimal source of information under varying missing data scenarios, guiding cross-modal attention and feature fusion for the final sentiment prediction based on the completed representations.

The main contributions of this work are:
\begin{itemize}
    \item We identify the challenge in MSA caused by the simultaneous occurrence of both intra-modality and inter-modality missingness with unknown patterns, and propose the \textbf{Senti-iFusion} framework as a solution.
    \item We introduce \textbf{Senti-iFusion}, an integrity-centered hierarchical fusion framework (\textbf{\textit{IE-IC-IF}}) that integrates \textit{Integrity Estimation, Integrity-weighted cross-modal Completion, and Integrity-guided multimodal Fusion}, introducing innovations in the learning paradigm, module interactions, and loss function design. 
    \item We validate the effectiveness of \textbf{Senti-iFusion} through experiments on publicly available MSA datasets, comparing it with existing methods, and demonstrate its advantage in fine-grained sentiment analysis tasks.
\end{itemize}

\section{Related Work}

\subsection{Context-based Multimodal Sentiment Analysis}

Early context-based MSA research assumed all modalities were fully available and focused on modeling rich cross-modal interactions. The field was initially advanced by resources and models that captured complex inter-modal dynamics: the CMU Multimodal SDK first unified datasets and baselines, while Tensor Fusion Network (TFN) \cite{zadeh2017TFN} captured unimodal, bimodal, and trimodal relations via computationally intensive tensor outer products; its successor, the Memory Fusion Network, enhanced temporal context modeling by incorporating a Delta-memory Attention mechanism. The rise of transformer architectures marked a significant shift, with models like MulT \cite{tsai2019multTransformer} utilizing directional cross-modal attention to achieve long-range, alignment-free fusion. Subsequently, MAG-BERT \cite{rahman2020MAG} demonstrated the effective injection of non-linguistic cues into pretrained language models via a dedicated Multimodal Adaptation Gate.Further extending this line of work, transformer-based architectures have continued to evolve; Key-Sparse Transformer \cite{chen2022keysparse} leveraged RoBERTa\cite{liu2019roberta} and Wav2vec\cite{schneider2019wav2vec} to focus on emotion-relevant information, enabling more efficient recognition, while TDFNet \cite{zhao2023tdfnet} introduced a deep-scale fusion mechanism to enhance multimodal emotional feature extraction through improved inter-modal interaction.

In parallel, substantial progress was made at the representation level through disentanglement and self-supervision. Representation-level advances\cite{yao2024drfuse,yang2022disentangled, selfmm2021, misa2020, kim2018disentangling} include MISA \cite{misa2020}, pioneered the learning of distinct modality-invariant and modality-specific representations in separate subspaces. Building on this, Self-MM \cite{selfmm2021} further leveraged self-supervised learning to generate unimodal pseudo-labels, thereby enforcing both cross-modal consistency and discrepancy. Other notable efforts \cite{kim2018disentangling, yang2022disentangled} have also explored various disentanglement strategies to obtain cleaner and more robust multimodal representations.
More recently, methods like DRFuse \cite{yao2024drfuse} have continued this trend by dynamically routing information through dual pathways to achieve effective modality fusion.
Although these approaches excel at contextual fusion via attention, memory, or disentanglement, their reliance on complete, well-aligned inputs makes them vulnerable to noisy or missing modalities.

\subsection{Noise-aware Multimodal Sentiment Analysis}

To enhance the robustness of MSA in real-world scenarios where data may be noisy or partially absent, researchers have developed methods that primarily fall into two categories: modality gating and completion/denoising.

Modality gating techniques dynamically assess the reliability of each modality and adjust their influence during fusion. A representative approach is MAG \cite{rahman2020MAG}, which omits a modality entirely when it is absent, allowing the model to depend on the remaining reliable cues. Language-guided models such as ALMT \cite{ALMT2023} extend this idea by dynamically emphasizing the modality richest in sentiment (typically language) and aligning other modalities to it, thereby maintaining robustness when auxiliary modalities are noisy or missing.

Completion and denoising techniques, on the other hand, aim to reconstruct missing or corrupted data rather than discarding it. This category includes both simple and advanced reconstruction strategies. Basic missing data imputation methods fill absent modalities with substitutes such as zero vectors \cite{parthasarathy2020training} or class-wise average values \cite{zhang2022deep}. More advanced missing data reconstruction approaches leverage generative models to synthesize plausible missing content. For instance, TFR-Net \cite{yuan2021TFR-Net} employs Transformer-based architectures to reconstruct missing segments within modality sequences. Other notable methods include using autoencoders for data imputation \cite{duan2014deep}, the Cascaded Residual Autoencoder (CRA) for reconstruction \cite{tran2017missing}, encoder-decoder networks for generating high-quality missing modality images \cite{cai2018deep}, and frameworks combining GANs with metric learning \cite{suo2019metric}. Additionally, adversarial training methods like NIAT \cite{yuan2023NIAT} and its extension \cite{yuan2023noise} align representations of clean and corrupted inputs to ensure robust sentiment encoding under noise.

By either reweighting reliable inputs or reconstructing missing ones, these approaches collectively enhance model resilience, supporting more robust sentiment analysis in practical applications with imperfect multimodal data.

\subsection{Representation Learning under Uncertain Missingness}

Learning effective multimodal representations under uncertain missing conditions remains a core challenge. Early approaches in this direction aimed to learn meaningful joint representations directly from incomplete data. For instance, Pham et al. \cite{pham2019found} proposed a cyclic translation framework to learn robust joint representations via cycle consistency. Following this, the MMIN model \cite{zhao2021missing} combined a Cascaded Residual Autoencoder (CRA) with cycle-consistent learning to imagine missing modalities and extract joint representations, establishing a mainstream paradigm for handling uncertain missingness.

Recent efforts have built upon this foundation, seeking to bridge both inter- and intra-modality gaps \cite{zhu2025proxy, chen2024cdrm,li2021noisyRobust,geng2021uncertainMultiview}. Unified architectures such as UMDF \cite{li2024UMDF} and LNLN \cite{LNLN2024} fuse cross- and intra-modal cues, yet they often rely on coarse, static estimates of modality completeness. Cross-modal generators like HRLF \cite{hrlf2024} and DrFuse \cite{yao2024drfuse} aim to rebuild missing modalities but generally lack integrity awareness, risking the amplification of noise during reconstruction. Prompt-based methods, such as MPLMM \cite{mplmm2024}, adapt flexibly to missing modalities but typically fail to account for the simultaneous loss of both intra- and inter-modality information, and are heavily dependent on large pretrained models.

In contrast, our proposed Senti-iFusion framework introduces continuous integrity scoring at both modality and segment levels, applies integrity-weighted completion to mitigate noisy reconstructions, and performs integrity-guided fusion that dynamically selects the most reliable dominant modality. This joint approach addresses two-level missingness more comprehensively and demonstrates superior robustness compared to prior works

\begin{figure*}[ht]
\centering
\includegraphics[width=1.0\textwidth]{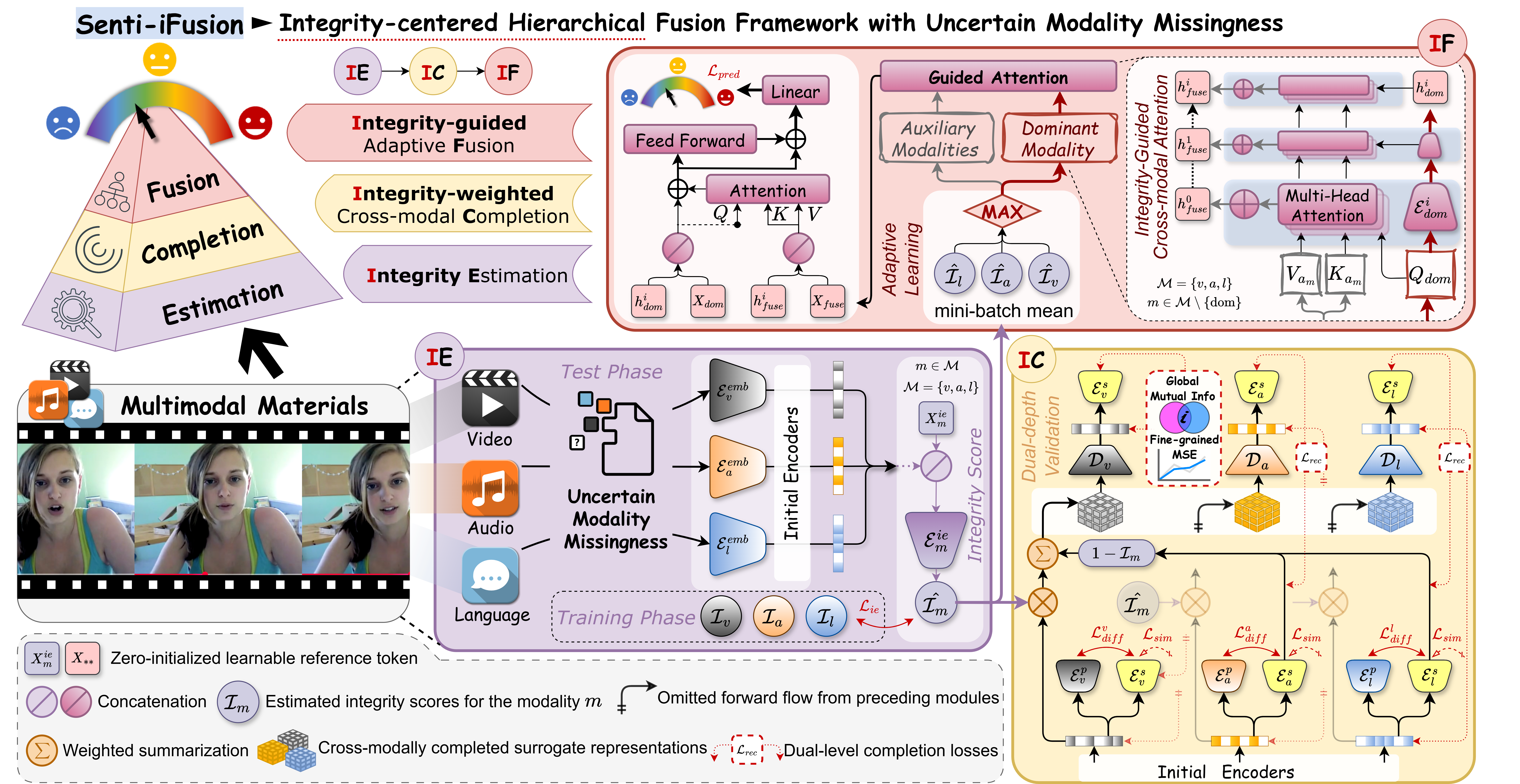} 
\caption{
Framework Overview. \textbf{Senti-iFusion} consists of three key components: \textbf{I}ntegrity \textbf{E}stimation (IE), \textbf{I}ntegrity-weighted Cross-modal \textbf{C}ompletion (IC), and \textbf{I}ntegrity-guided Adaptive \textbf{F}usion (IF).
Given multimodal inputs with \emph{unknown and mixed missingness}, the model first estimates the integrity scores of each modality within the mini-batch.  
The predicted scores guide the \textit{integrity-weighted} cross-modal feature completion,  using dual-depth validation and a dual-level loss (semantic and fine-grained).  
Finally, the \textit{integrity-guided} attention mechanism adaptively cooperates with the dominant modality and performs multi-scale fusion for sentiment prediction.
}
\label{fig:framework}
\end{figure*}

\begin{table}[t]
\centering
\caption{Definition of Notations}
\label{tab:notation}
\begin{tabular}{ll}
\toprule
\textbf{Notation} & \textbf{Definition} \\
\midrule
$m$ & Modality index ($m \in \{l, a, v\}$ for language, acoustic, visual). \\
$U_{m}$ & Original input features of modality $m$. \\
$\tilde{U}_m$ & Incomplete features of modality $m$ after random missingness. \\
$\mathcal{E}_m^{\mathrm{emb}}$ & Embedding encoder for modality $m$. \\
$\hat{u}_m$ & Embeddings corresponding to $\tilde{U}_m$. \\
$X$ & Zero-initialized extra token (e.g., $X_m^{\mathrm{emb}}$, $X^{\mathrm{ie}}$). \\
$\mathcal{I}_m$ & Ground-truth integrity score of modality $m$. \\
$\mathcal{E}_m^{\mathrm{ie}}$ & Integrity estimation encoder for modality $m$. \\
$\hat{\mathcal{I}}_m$ & Predicted integrity score of modality $m$. \\
$\mathcal{E}_m^{s}$ & Modality-invariant (shared) encoder for modality $m$. \\
$\mathcal{E}_m^{p}$ & Modality-specific (private) encoder for modality $m$. \\
$h_m^{\mathrm{s}}, h_m^{\mathrm{p}}$ & Shared and private hidden features of modality $m$. \\
$h_m^{\mathrm{sur}}$ & Surrogate features of modality $m$. \\
$\mathcal{D}_m$ & Decoder for modality $m$. \\
$\tilde{u}_m$ & Reconstructed features of modality $m$. \\
$\mathcal{E}_{\mathrm{dom}}^i$ & $i$-th layer encoder for dominant modality processing. \\
$h_{\mathrm{dom}}^i$ & $i$-th layer dominant-modality features. \\
$A$ & Auxiliary-modality features. \\
$h_{\mathrm{fuse}}^i$ & $i$-th layer fused multimodal features. \\
$\mathcal{L}_{\mathrm{rec}}$ & Reconstruction loss. \\
$\mathcal{L}_{\mathrm{ie}}$ & Integrity estimation loss. \\
$\mathcal{L}_{\mathrm{mse}}$ & Mean squared error loss. \\
$\mathcal{L}_{\mathrm{mi}}$ & Mutual information loss. \\
$\mathcal{L}_{\mathrm{pred}}$ & Prediction loss. \\
$\lambda_{mse}^t, \lambda_{mi}^t$ & Balance factors of $\mathcal{L}_{\mathrm{mse}}$ and $\mathcal{L}_{\mathrm{mi}}$ for scale $t$. \\
$\alpha, \beta, \sigma$ & Balance factors of $\mathcal{L}_{\mathrm{ie}}$, $\mathcal{L}_{\mathrm{rec}}$, and $\mathcal{L}_{\mathrm{pred}}$. \\
\bottomrule
\end{tabular}
\end{table}

\section{Methodology}

\subsection{Problem Definition}

We focus on multimodal sentiment analysis under scenarios of completely unknown missingness. Each video in the dataset is segmented into individual utterances, each consisting of three modality-specific sequences: language ($l$), acoustic ($a$), and visual ($v$). Let \(\mathcal{M} = \{l, a, v\}\). Formally, each utterance is denoted as \(U=[U_l, U_a, U_v]\) , where \(U_m \in \mathbb{R}^{T_m \times d_m}\) with \(m \in \mathcal{M}\), \(T_m\) representing the sequence length, and \(d_m\) denoting the corresponding feature dimension.
To simulate real-world scenarios, we define two concurrent missingness conditions: (1) \textit{inter-modality missingness}, where one or more modality sequences may be absent; and (2) \textit{intra-modality missingness}, where individual tokens within a modality sequence are missing. Our framework incorporates both types of missingness to reflect realistic multimodal data corruption.

\subsection{Data Preparation}

\subsubsection{Multimodal Input Processing}
Consistent with prior works \cite{yu2021learning}, we use BERT to encode language data, Librosa to extract acoustic features, and OpenFace \cite{baltruvsaitis2016openface} to obtain visual features. The pre-processed sequences are represented as \(U = [U_l, U_a, U_v]\), which are then passed through a random data missingness procedure to simulate incomplete data.

\subsubsection{Random Data Missing Simulation}
Building on previous approaches \cite{mplmm2024, yuan2021TFR-Net}, we apply two levels of missingness to simulate real-world scenarios. For modality-level missingness, entire modality sequences are randomly omitted. For intra-modality missingness, we randomly erase varying proportions of information (ranging from 0\% to 100\%) for each modality. Specifically, for visual and acoustic modalities, erased information is replaced with zeros, while for the language modality, missing data is replaced with \texttt{[UNK]}, representing an unknown word in BERT \cite{devlin2019bert}. The incomplete features resulting from this stochastic missingness simulation are denoted as \(\tilde{U} = [\tilde{U}_l, \tilde{U}_a, \tilde{U}_v]\).

\subsection{Integrity Estimation}

\subsubsection{Data Embedding}

To further extract and unify representations across modalities, we employ an initial encoder \(\mathcal{E}_{emb}^m\), consisting of a linear projection layer followed by a two-layer Transformer. For each modality \(m \in \mathcal{M}\), zero-initialized extra tokens \(X_m^{emb}\) are prepended to the incomplete multimodal inputs \(\tilde{U}_m\). These tokens serve as aggregation anchors that interact with all time steps via attention enabling the Transformer encoder to learn a global predictive representation without relying on manual pooling strategies. The concatenated sequence is then fed into \(\mathcal{E}_{emb}^m\) to yield the unified representation \(\hat{u}_m \). The process is defined as:

\begin{equation}\label{eq:modality_embedding}
 \hat{u}_m = \mathcal{E}_m^{emb}(\text{Concat}(X_m^{emb}, \tilde{U}_m)) \in \mathbb{R}^{T \times d}
\end{equation}

\subsubsection{Integrity Prediction}
To estimate the extent of missingness within each modality sequence, we apply an encoder \(\mathcal{E}_{ie}^m\), consisting of a two-layer Transformer followed by a linear projection layer to produce a continuous integrity score. During training, ground-truth integrity labels \(\mathcal{I}_m\) are computed as the complement of the missing ratio applied during intra-modality random masking. For instance, a modality sequence with 60\% missing yields an integrity score of 0.4. The estimation process could be described as:

\begin{equation}\label{eq:integrity_estimation}
 \hat{\mathcal{I}_m} = \mathcal{E}_{m}^{ie}(\text{Concat}(X_{ie}, \hat{u}_m))
\end{equation}
where \(X_{ie} \in \mathbb{R}^{T \times d}\) is a zero-initialized extra token prepended to the input, enabling the Transformer to capture missing patterns and predict integrity score \(\hat{\mathcal{I}}\). The process is optimized utilizing Mean Square Error (MSE) loss:

\begin{equation}\label{eq:integrity_loss}
 \mathcal{L}_{ie} = \frac{1}{N} \sum_{k=0}^{N} \| \hat{\mathcal{I}}^k - \mathcal{I}^k \|_2^2
\end{equation}

where \(N\) is the number of training samples, and \(\hat{\mathcal{I}}^k\) denotes the predicted integrity score for the \(k\)-th sample.

\subsection{Integrity-weighted Cross-modal Completion}

\subsubsection{Surrogate Representation Completion}

Incomplete or noisy modality sequences introduce ambiguity in sentimental cues and cause redundancy during fusion, impeding the extraction of valuable information. Prior work \cite{hrlf2024} addressed this via feature disentanglement, separating sentiment-relevant and modality-specific components to recover missing tokens and enhance modality-level interaction. However, due to distributional shifts introduced by deep encoders, such decoders often struggle to accurately reconstruct features for downstream tasks.

To mitigate this, we incorporate an integrity-guided weighting mechanism to generate surrogate hidden representations that integrate both modality-invariant semantic structures and modality-specific features from language, acoustic, and visual inputs. Specifically, we employ two parallel Transformer encoders $\mathcal{E}_m^s$ and $\mathcal{E}_m^p$ to extract disentangled representations:

\begin{equation}\label{eq:emotional_semantics}
\hat{h}_m^s = \mathcal{E}_m^s(\hat{u}_m), \quad
\hat{h}_m^p = \mathcal{E}_m^p(\hat{u}_m), \quad
\hat{h}_m^s, \hat{h}_m^p \in \mathbb{R}^{T \times d}
\end{equation}

Here, superscripts $s$ and $p$ indicate the shared semantics structures and private modality-specific representations, respectively. To enforce the semantic alignment across modalities, we apply a similarity loss $\mathcal{L}_{sim}$, encouraging $\hat{h}_m^s$ from different modalities to remain close:

\begin{equation}\label{eq:similarity_loss}
 \mathcal{L}_{sim} = \sum_{(x, y) \in \mathcal{P}} \frac{1}{N} \| \hat{h}_x^s - \hat{h}_y^s \|_2^2
\end{equation}
where $\mathcal{P} = \{(l,a), (l,v), (a,v)\}$ denotes all cross-modal modality pairs. This formulation encourages semantic consistency across modalities while preserving modality-specific characteristics.

To ensure representation disentanglement within each modality, we introduce a difference loss $\mathcal{L}_{diff}^m$, which penalizes similarity between shared ($\hat{h}_m^s$) and private ($\hat{h}_m^p$) representations:

\begin{equation}\label{eq:difference_loss}
\mathcal{L}_{diff}
= \frac{1}{|\mathcal{M}|} \sum_{m \in \mathcal{M}}
    \frac{1}{d_m^2}
    \left\|
        \big(\tilde{h}_m^s\big)^{\top} \tilde{h}_m^p
    \right\|_F^2
\end{equation}
where $\tilde{h}_m^s$ and $\tilde{h}_m^p$ denote the zero-mean and $L_2$-normalized shared and private representations for modality $m$, respectively.

This correlation-based loss penalizes the squared Frobenius norm of the cross-covariance between 
$\tilde{h}_m^s$ and $\tilde{h}_m^p$, enforcing orthogonality and promoting representational independence within each modality.

Thus, total loss optimizing the encoder part could be formulated as:
\begin{equation}\label{eq:loss_enc}
  \mathcal{L}_{rec}^{enc} = \sum_{(x, y) \in \mathcal{P}} \mathcal{L}_{sim}^{xy} 
  + \sum_{m \in \mathcal{M}} \mathcal{L}_{diff}^m
\end{equation}

Then, to leverage complementary cues across modalities in the presence of missing data, we construct a surrogate feature \(\hat{h}_m^{sur}\) that adaptively combines the original input and cross-modal recovery based on the estimated integrity score:

\begin{equation}\label{eq:surrogate_generation}
 \hat{h}_m^{sur} = \hat{\mathcal{I}}_m \cdot (\hat{u}_m) + (1 - \hat{\mathcal{I}}_m) \cdot (\hat{h}_{m_1}^{s_1} + \hat{h}_{m_2}^{s_2})
\end{equation}
Here, \(\hat{h}_{m_1}\) and \(\hat{h}_{m_2}\) denote the shared semantic representations extracted from the other two modalities. This design enables the model to selectively rely on recovered semantics when the integrity of the current modality is low.

\subsubsection{Dual-Depth Validation with Dual-level Completion Loss}
To preserve semantic information and maintain low-level distributional fidelity, we introduce a dual-depth validation scheme with dual-level completion loss. 
The dual-depth validation involves two representations: the reconstructed embeddings decoded by the Transformer Decoder \(\mathcal{D_{\text{m}}}\) based on the surrogate representations, and the sentiment semantic features derived from the reconstructed embeddings via the Shared Encoder \(\mathcal{E}_m^s\). The dual-level loss consists of MSE loss and Mutual Information (MI) loss. To preserve the semantic structure, we apply MI loss following Deep InfoMax \cite{hjelm2018MILoss} to quantify the semantic information, as described below: 

\begin{equation}\label{eq:rec_global&semantics1}
  I(X;Y) = \mathbb{E}_{p(x,y)}\!\left[\log\frac{p(x,y)}{p(x)p(y)}\right]
\end{equation}

\begin{equation}\label{eq:rec_global&semantics2}
    \begin{split}        
      I(X;Y) \ge \mathbb{E}_{p(x,y)}\left[-\log\left(1 + e^{-T_\theta(x,y)}\right)\right] \\
       + \mathbb{E}_{p(x)p(y)}\left[-\log\left(1 + e^{T_\theta(x,y)}\right)\right]  
    \end{split}
\end{equation}
where \(T_\theta(x, y)\) is a discriminator network designed for lower-bound approximation. This formulation encourages representations \(x\) and \(y\) to be mutually predictive when sampled from the joint distribution, and non-informative otherwise.

Then we optimize it as a component of total loss to encourage semantic-level alignment as follows:

\begin{equation}\label{eq:rec_global&semantics}
  \tilde{u}_{m} = \mathcal{D}_m(\hat{h}_m^{sur}),\quad
  \tilde{h}_{m}^s = \mathcal{E}_m^s(u_{m}^{rec}),\quad
  \tilde{u}_{m}, \tilde{h}_{m}^s \in \mathbb{R}^{T \times d}
\end{equation}

\begin{equation}\label{eq:mi_global&semantics}
 \mathcal{L}_{mi}^{g} = - \sum^m MI(\tilde{u}_{m}, u_m),\quad
 \mathcal{L}_{mi}^{s} = - \sum^m MI(\tilde{h}_{m}, h_m^s)
\end{equation}
where $u_m$ and \(h_m^s\) denote the original complete inputs and their corresponding semantic features from the encoder, respectively. This loss promotes accurate feature restoration across both global and semantic levels.

while MSE constrains fine-grained reconstruction accuracy, which is defined as follows:

\begin{equation}\label{eq:mse_global}
  \mathcal{L}_{mse}^{g} = \frac{1}{N} \sum^m \| \tilde{u}_{m} - u_m \|_2^2,\quad
  \mathcal{L}_{mse}^{s} = \frac{1}{N} \sum^m \| \tilde{h}_{m} - h_m^s \|_2^2
\end{equation}

The overall completion loss for the decoder can then be formulated as:

\begin{equation}\label{eq:loss_dec}
  \mathcal{L}_{rec}^{dec} = \sum_{t \in \{g, s\}} \left( \lambda_{mse}^t \cdot \mathcal{L}_{mse}^t + \lambda_{mi}^t \cdot \mathcal{L}_{mi}^t \right)
\end{equation}
where $\lambda_{mse}^t$ and $\lambda_{mi}^t$ are coefficients for corresponding loss.

\subsection{Integrity-guided Adaptive Modality Fusion}
\subsubsection{Modality Fusion}

Unlike prior approaches that arbitrarily fix language as the dominant modality, we propose a data-driven fusion strategy that adaptively selects the most informative modality based on predicted integrity scores and cross-modal reconstruction quality. Specifically, given modality-wise integrity weights \( w_l \), \( w_a \), and \( w_v \), we compute the average integrity score across the batch and select the modality with the highest mean score as the dominant one:

\begin{equation}\label{eq:dom_chosen}
  \text{dom} = \arg \max \left( \sum_{i=1}^{N} \mathbb{I}(\text{dom}_i = \text{modality}) \right)
\end{equation}
let $dom$ denotes the selected dominant modality, and $a_1, a_2 \in \mathcal{M} \setminus \{\text{dom}\}$ represent the auxiliary modalities.

To progressively refine dominant features, we use a multi-layer Transformer encoder:

\begin{equation}\label{eq:adaptive_dom}
  h_{dom}^{i} = \mathcal{E}_{dom}^{i}(h_{dom}^{i-1}) \in \mathbb{R}^{T \times d}
\end{equation}
where \(i \in \{2, 3\}\) denotes the \(i\)-th layer of Transformer encoders.

Inspired by \cite{ALMT2023}, we integrate auxiliary modalities via cross-modal attention, using the dominant features \(h_{dom}^i \) as query and auxiliary features \(A_1, A_2\) as key and value. Let \( m \in \mathcal{M} \setminus \{\text{dom}\} \) denote the auxiliary modalities. For each fusion layer \( j \in \{1, 2, 3\} \), the fusion feature is updated as:

\begin{equation}\label{eq:fusion_general}
  h_{\text{fuse}}^{j} = h_{\text{fuse}}^{j-1} + \sum_{m \in \mathcal{M} \setminus \{\text{dom}\}} \gamma_{m} V_{m}
\end{equation}

For each auxiliary modality $a_{m}$, the similarity matrix is computed as:

\begin{equation}\label{eq:attn_general}
  \gamma_{m} = \mathrm{softmax}\left( \frac{Q_{\text{dom}} K_{m}^\top}{\sqrt{d_k}} \right)
\end{equation}
with \( Q_{\text{dom}} = h_{\text{dom}}^j W_Q \), \( K_{m} = A_{m} W_K^m \), and \( V_{m} = A_{\mu} W_V^m \). Here \( W_Q, W_K^m, W_V^m \in \mathbb{R}^{d \times d_k} \) are learnable projection matrices.

\subsubsection{Prediction}

To perform sentiment prediction, we utilize the fused feature \(h_{fuse}^{3}\), which integrates multi-modal and multi-scale information. To further enhance interaction between the fused and dominant modality representations, we concatenate zero-initialized tokens $X_{dom}$ and $X_{fuse}$ with $h_{dom}^{3}$ and $h_{fuse}^{3}$, respectively:

\begin{equation}\label{eq:H_dom}
\begin{aligned}
  H_{dom}  &= \text{Concat}(X_{dom},\; h_{dom}^{3}), \\
  H_{fuse} &= \text{Concat}(X_{fuse},\; h_{fuse}^{3}), \\
\end{aligned}
\end{equation}
where $H_{dom},\; H_{fuse} \in \mathbb{R}^{(T+1) \times d}$. The two are then passed into a cross-modal Transformer encoder $\mathcal{E}_{pred}$ to produce the final predictive representation:

\begin{equation}\label{eq:pred_feat}
  h_{pred} = \mathcal{E}_{pred}(H_{dom}, H_{fuse}) \in \mathbb{R}^{1 \times d}
\end{equation}

Finally, $h_{pred}$ is fed into a linear prediction head to generate the output $\hat{y}$. We optimize the prediction using MSE loss:

\begin{equation}\label{eq:pred_loss}
  \mathcal{L}_{pred} = \frac{1}{N} \sum_{k=0}^{N} \| \hat{y}^k - y^k \|_2^2
\end{equation}

\subsection{Overall Learning Objectives}

To summarize, the overall learning objective of our model combines three components: (1) Integrity Estimation loss $\mathcal{L}_{ie}$, calculated as the sum of MSE losses across language, acoustic and visual modalities; (2) Integrity-weighted Cross-modal Completion loss $\mathcal{L}_{rec}$, including both encoder- and decoder- level completion constraints; and (3) Prediction loss $\mathcal{L}_{pred}$ for Integrity-guided Adaptive Modality Fusion module.

\begin{equation}
\label{eq:loss_total}
\mathcal{L}_{total} =\\
\begin{cases}
\alpha \mathcal{L}_{ie} + \beta \mathcal{L}_{rec}, & \text{stage 1},\\
\alpha \mathcal{L}_{ie} + \beta \mathcal{L}_{rec} + \sigma \mathcal{L}_{pred}, & \text{stage 2}.
\end{cases}
\end{equation}
Here $\alpha, \beta, \sigma$ are hyperparameters that balance the losses, respectively.

The training procedure for our Integrity Estimation module is performed in two progressive stages to ensure stable convergence. Initially, the model learns to predict modality completeness from input representations. In this phase, we freeze all model parameters except the projection layers and the integrity estimation module. The prediction loss is not included at this stage. Subsequently, in the second stage, the estimated integrity scores guide the cross-modal completion and fusion training. Formally, our total loss is defined in Equation \ref{eq:loss_total}.

\begin{table*}[ht]
\small
\centering
\caption{Performance comparison of Senti-iFusion with baseline models on \textsc{MOSI} and \textsc{MOSEI} under 50\% inter- and intra-modality  missingness. All models are trained with identical seeds to ensure reproducibility.}

\label{tab:metric_comparison_drop0.5}
\begin{tabular}{c c c c c c c c c c c c}
\hline
\multirow{2}{*}{\textbf{Model}}
& \multicolumn{5}{c}{\textbf{MOSI}} 
& \multicolumn{5}{c}{\textbf{MOSEI}} \\
\cline{2-6} 
\cline{8-12}
& MAE($\downarrow$) & Acc-7($\uparrow$) & Acc-5($\uparrow$) & Acc-2($\uparrow$) & F1($\uparrow$) &  
& MAE($\downarrow$) & Acc-7($\uparrow$) & Acc-5($\uparrow$) & Acc-2($\uparrow$) & F1($\uparrow$) \\
\hline
MISA (MM2020) & 1.2753 & 0.2522 & 0.2536 & 0.6152 & \underline{0.6442} & & 0.7264 & 0.4385 & 0.4443 & 0.7254 & \textbf{0.8091} \\
Self-MM (AAAI2021) & 1.2557 & 0.2653 & 0.2843 & 0.5808 & 0.5969 & & 0.7304 & \underline{0.4480} & \underline{0.4531} &  0.7155 & 0.6810 \\
CubeMLP (MM2022) & 1.2687 & 0.2303 & 0.2391 & 0.6341 & 0.6243 & & \textbf{0.6910} & 0.3528 & 0.4315 &  0.7131 & 0.7187 \\
DMD (CVPR2023) & 1.2825 & 0.2099 & 0.2143 & 0.6128 & 0.6151 & & 0.7330 & \underline{0.4480} & 0.4520 &  0.7066 & 0.7099 \\
LNLN (NeurlPS2024) & \underline{1.1925} & \underline{0.2915} & \underline{0.3120} & \underline{0.6723} & 0.6405 & & 0.7247 & 0.4456 & 0.4516 & \underline{0.7262} & 0.7040 \\
\textbf{Senti-iFusion}   & \textbf{1.1595} & \textbf{0.3047} & \textbf{0.3324} & \textbf{0.6784} & \textbf{0.6842} & & \underline{0.6959} & \textbf{0.4507} & \textbf{0.4559} & \textbf{0.7276} & \underline{0.7406} \\
\hline
\end{tabular}
\end{table*}

\begin{table*}[t]
\centering
\small
\caption{Performance comparison on \textsc{MOSI} under complete inter-modality corruption ($drop\_rate = 1.0$), evaluated across six modality-retention Modes (0--5). Each Mode preserves a specific subset of modalities while completely masking the others.}
\label{tab:mode_all}

\begin{adjustbox}{max width=\textwidth}
\begin{tabular}{c l l c c c c c}
\toprule
\textbf{Mode} & \textbf{Available Modalities} & \textbf{Model} &
\textbf{MAE$\downarrow$} & \textbf{Acc-7$\uparrow$} &
\textbf{Acc-5$\uparrow$} & \textbf{Acc-2$\uparrow$} &
\textbf{F1$\uparrow$} \\
\midrule

\multirow{6}{*}{0} & \multirow{6}{*}{\{a, v\}} 
& MISA            & 1.4478 & 0.1545 & 0.1545 & 0.4548 & 0.5917 \\
& & Self-MM        & 1.4315 & 0.1589 & 0.1589 & 0.5351 & 0.5377 \\
& & CubeMLP        & 1.5302 & 0.1720 & 0.1764 & 0.5160 & 0.5137 \\
& & DMD            & 1.5216 & 0.1516 & 0.1516 & 0.4223 & 0.2507 \\
& & LNLN           & 1.3654 & 0.2114 & 0.2114 & 0.6052 & 0.5669 \\
& & \textbf{Senti-iFusion}  
                   & \textbf{1.3327} & \textbf{0.1953} & \textbf{0.1983} & \textbf{0.5930} & \textbf{0.6180} \\
\midrule

\multirow{6}{*}{1} & \multirow{6}{*}{\{l, v\}}
& MISA            & 1.2872 & 0.2085 & 0.2099 & 0.6268 & 0.6594 \\
& & Self-MM        & 1.1329 & 0.3178 & 0.3455 & 0.6951 & 0.6968 \\
& & CubeMLP        & 1.2242 & 0.2828 & 0.3090 & 0.6327 & 0.6337 \\
& & DMD            & 1.1708 & 0.2711 & 0.2813 & 0.6738 & 0.6753 \\
& & LNLN           & 1.1192 & 0.3149 & 0.3411 & 0.7104 & 0.7116 \\
& & \textbf{Senti-iFusion}  
                   & \textbf{1.0972} & \textbf{0.3265} & \textbf{0.3630} & \textbf{0.7210} & \textbf{0.7319} \\
\midrule

\multirow{6}{*}{2} & \multirow{6}{*}{\{l, a\}}
& MISA            & 1.1887 & 0.2595 & 0.2741 & 0.6764 & 0.6486 \\
& & Self-MM        & 1.1653 & 0.2872 & 0.3134 & 0.6662 & 0.6661 \\
& & CubeMLP        & 1.3117 & 0.2376 & 0.2595 & 0.6458 & 0.6393 \\
& & DMD            & 1.2089 & 0.2857 & 0.3003 & 0.6738 & 0.6741 \\
& & LNLN           & 1.1188 & 0.3280 & 0.3644 & 0.7180 & 0.7184 \\
& & \textbf{Senti-iFusion}  
                   & \textbf{1.0906} & \textbf{0.3397} & \textbf{0.3776} & \textbf{0.7104} & \textbf{0.7110} \\
\midrule

\multirow{6}{*}{3} & \multirow{6}{*}{\{v\}}
& MISA            & 1.4498 & 0.1545 & 0.1545 & 0.4548 & 0.5944 \\
& & Self-MM        & 1.4306 & 0.1545 & 0.1545 & 0.4223 & 0.2507 \\
& & CubeMLP        & 1.5591 & 0.1866 & 0.1983 & 0.5277 & 0.5265 \\
& & DMD            & 1.4518 & 0.1574 & 0.1574 & 0.4695 & 0.3963 \\
& & LNLN           & 1.3842 & 0.2114 & 0.2945 & 0.5777 & 0.4231 \\
& & \textbf{Senti-iFusion}  
                   & \textbf{1.3837} & \textbf{0.1866} & \textbf{0.1866} & \textbf{0.5762} & \textbf{0.5736} \\
\midrule

\multirow{6}{*}{4} & \multirow{6}{*}{\{a\}}
& MISA            & 1.4481 & 0.1545 & 0.1545 & 0.4519 & 0.5877 \\
& & Self-MM        & 1.4505 & 0.1589 & 0.1589 & 0.4863 & 0.4119 \\
& & CubeMLP        & 1.5223 & 0.1633 & 0.1676 & 0.4475 & 0.2767 \\
& & DMD            & 1.5279 & 0.1516 & 0.1516 & 0.4223 & 0.2507 \\
& & LNLN           & 1.3857 & 0.2114 & 0.2114 & 0.5777 & 0.4231 \\
& & \textbf{Senti-iFusion}  
                   & \textbf{1.3624} & \textbf{0.1939} & \textbf{0.1939} & \textbf{0.5991} & \textbf{0.6008} \\
\midrule

\multirow{6}{*}{5} & \multirow{6}{*}{\{l\}}
& MISA            & 1.1813 & 0.2843 & 0.2974 & 0.6647 & 0.6537 \\
& & Self-MM        & 1.1275 & 0.2959 & 0.3222 & 0.6875 & 0.6868 \\
& & CubeMLP        & 1.1638 & 0.2799 & 0.3222 & 0.6662 & 0.6667 \\
& & DMD            & 1.2420 & 0.2653 & 0.2770 & 0.6585 & 0.6590 \\
& & LNLN           & 1.1110 & 0.3338 & 0.3557 & 0.7088 & 0.7101 \\
& & \textbf{Senti-iFusion}  
                   & \textbf{1.1052} & \textbf{0.3251} & \textbf{0.3601} & \textbf{0.7012} & \textbf{0.7068} \\
\bottomrule

\end{tabular}
\end{adjustbox}
\end{table*}

\section{Experiments}

\subsection{Experiment Settings}
This section summarizes the experimental setup, including datasets, baselines, evaluation metrics, missing data patterns, and implementation details.

\begin{figure}[ht]
\centering
\includegraphics[width=1.0\columnwidth]{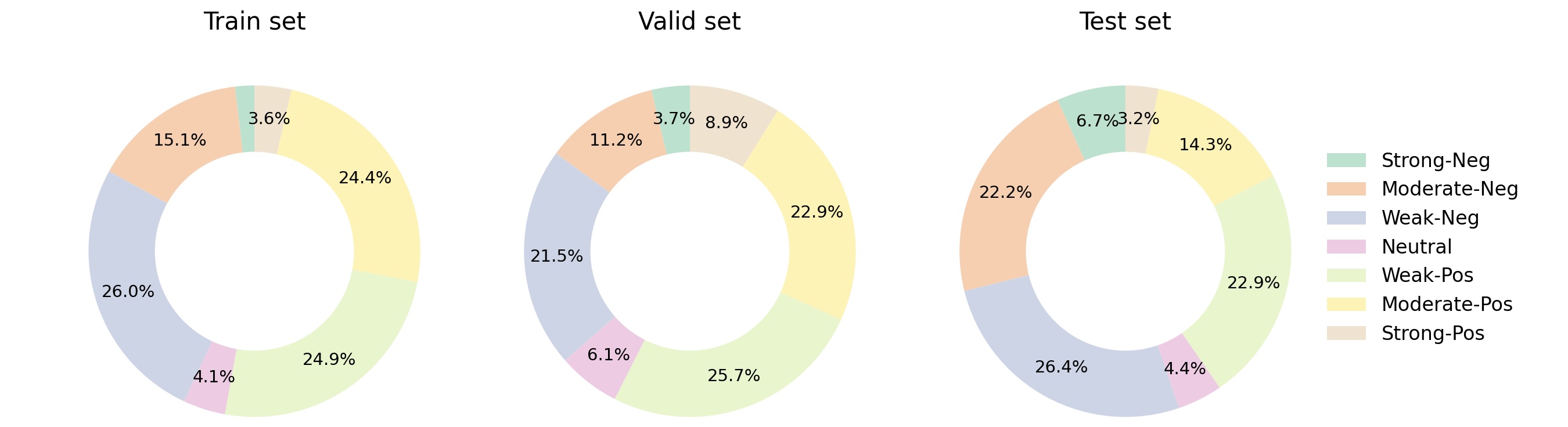} 
\caption{
Sentiment score distributions of the MOSI dataset across the training, validation, and test splits, illustrating label imbalance and distribution shifts between splits.
}
\label{fig: mosi_distribution}
\end{figure}

\begin{figure}[ht]
\centering
\includegraphics[width=1.0\columnwidth]{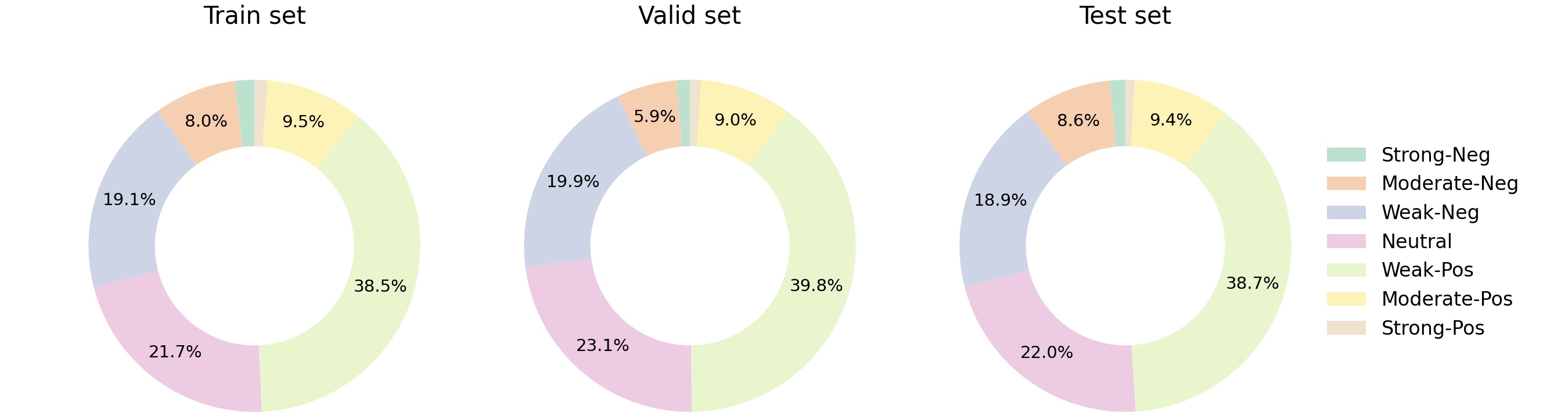} 
\caption{
Sentiment score distributions of the MOSEI dataset across the training, validation, and test splits, highlighting the pronounced class imbalance in real-world sentiment data.
}
\label{fig: mosei_distribution}
\end{figure}

\subsubsection{Datasets and Metrics}

We evaluate Senti-iFusion on two widely-used MSA benchmarks to simulate real-world scenarios.
\textbf{MOSI} \cite{zadeh2016mosi} consists of 2,199 video segments from 93 YouTube movie reviews, with splits of 1,284 for training, 229 for validation, and 686 for testing.
\textbf{MOSEI} \cite{zadeh2018MOSEI} extends MOSI, containing 22,856 segments with a split of 16,326 for training, 1,871 for validation, and 4,659 for testing. It offers a more diverse test for multimodal sentiment modeling.

Figure~\ref{fig: mosi_distribution} and Figure~\ref{fig: mosei_distribution} illustrate the sentiment score distributions for the MOSI and MOSEI datasets across training, validation, and test splits. These distributions reflect the inherent class imbalance and polarity bias present in real-world sentiment data. Such insights are critical for understanding model performance under diverse conditions.

\begin{itemize}
  \item \textbf{Category Imbalance}: Both data sets exhibit an evident class imbalance. For instance, in MOSEI (Figure~\ref{fig: mosi_distribution}), the category \textit{Weak-Positive} dominates, comprising approximately 38\% to 40\% of all splits.
  \item \textbf{Distribution Shift Between Splits}: MOSI shows nonuniform distributions across train/val/test. For example, \textit{Weak-Negative} samples account for 15.1\% in the training set, but 22.2\% in the test set (Figure~\ref{fig: mosei_distribution}).
  \item \textbf{Impact on Learning Under Missing Conditions}: The presence of unbalanced distributions may introduce bias, especially when modalities are missing. In such noisy settings, models may exhibit lazy learning behavior, favoring dominant sentiment categories instead of learning subtle distinctions.
\end{itemize}

These observations suggest that robust models must not only handle multimodal fusion but also generalize across inconsistent and biased data distributions.

Moreover, each sample in both datasets is labeled with a sentiment score from \(-3\) (highly negative) to \(+3\) (highly positive). We evaluate model performance using mean absolute error (MAE), 7-class accuracy (ACC-7), 5-class accuracy (ACC-5), 2-class accuracy on non-neutral samples (Non0\_ACC-2), and the corresponding F1 score.

\subsubsection{Baselines}
We compare \textbf{Senti-iFusion} against five strong baselines under the same data-missing conditions.

\textbf{MISA}\cite{misa2020}: A joint-learning model disentangles modality-specific and invariant representations to reduce modality gaps and enhance fusion.  

\textbf{Self-MM}\cite{selfmm2021}: A self-supervised approach generates unimodal pseudo-labels and disentangles features without human annotations.  

\textbf{CubeMLP}\cite{cubemlp2022}: A lightweight MLP-based fusion framework exploring cross-modal interactions.

\textbf{DMD}\cite{dmd2023}: A model handles modality heterogeneity through disentangled representation learning and dynamic cross-modal distillation via graphs. 

\textbf{LNLN}\cite{LNLN2024}: A method addresses intra-modality incompleteness using language as the dominant modality for fusion guidance.

All methods are evaluated under identical missing-modality conditions for fair comparison.

\subsubsection{Missing Pattern Settings}

We assume consistent missing patterns across training, validation, and testing. Each sample undergoes two types of missingness:  
 
(1) \textit{Inter-modality missingness}, where one or more modality sequences are randomly omitted, with at least one modality present;  

(2) \textit{Intra-modality missingness}, where tokens within a modality sequence are randomly masked. For modalities marked missing in the inter-modality step, their missing-rate list is set to all ones, simulating complete absence.

\subsubsection{Implementation Details}

\begin{table}[t]
\centering
\small
\caption{Hyperparameters used for training \textbf{Senti-iFusion} on MOSI and MOSEI.}
\label{tab:hyperparams}
\begin{tabular}{p{3.0cm} p{5.0cm}}
\toprule
\textbf{Category} & \textbf{Hyperparameters} \\
\midrule

\textbf{General Setup} &
Seed = 1112 \\
& Batch size = 64 \\
& Epochs = 150 (MOSI), 200 (MOSEI) \\
& Learning rate = $1\times10^{-4}$ \\
& Weight decay = $1\times10^{-4}$ \\
\midrule

\textbf{Loss Weights} &
$\lambda_{\text{corr}} = 0.1$ \\
& $\lambda_{\text{ccc}} = 0.05$ \\
& $\lambda_{\text{cross-mse}} = 0.5$ \\
& $\lambda_{\text{cross-mi}} = 0.4$ \\
& $\lambda_{\text{info-mse}} = 0.3$ \\
& $\lambda_{\text{info-mi}} = 0.2$ \\
\bottomrule
\end{tabular}
\end{table}
 
Senti-iFusion is implemented using PyTorch~2.4.0 on a server with an Intel\textsuperscript{\textregistered} Xeon\textsuperscript{\textregistered} Gold~6132 CPU and six NVIDIA GeForce RTX~3090 GPUs (24\, GB).  
Experiments are performed on unaligned MOSI and MOSEI sequences with a fixed random seed of 1112 for reproducibility.  
We set the input length $T=8$, hidden dimension $d=128$, and batch size to 64. The AdamW optimizer is used with an initial learning rate of $1\times10^{-4}$.  
Total loss weights are \((\alpha, \beta, \sigma) = (0.9, 0.4, 1.0)\), and decoder sub-loss weights are \(\lambda_{mse}^{g}=0.5\), \(\lambda_{mi}^{g}=0.4\), \(\lambda_{mse}^{s}=0.3\), and \(\lambda_{mi}^{s}=0.2\).  
The model is trained for 150 epochs, with the first 40 dedicated to integrity estimation (Stage 1), followed by joint training (Stage 2). A warm-up stage, a cosine annealing learning rate schedule, and early stopping are applied throughout training.  
Baseline results are reproduced using official implementations and default settings, with a unified data-missing pattern for fair comparison.

\begin{figure*}[t]
\centering
\includegraphics[width=\textwidth]{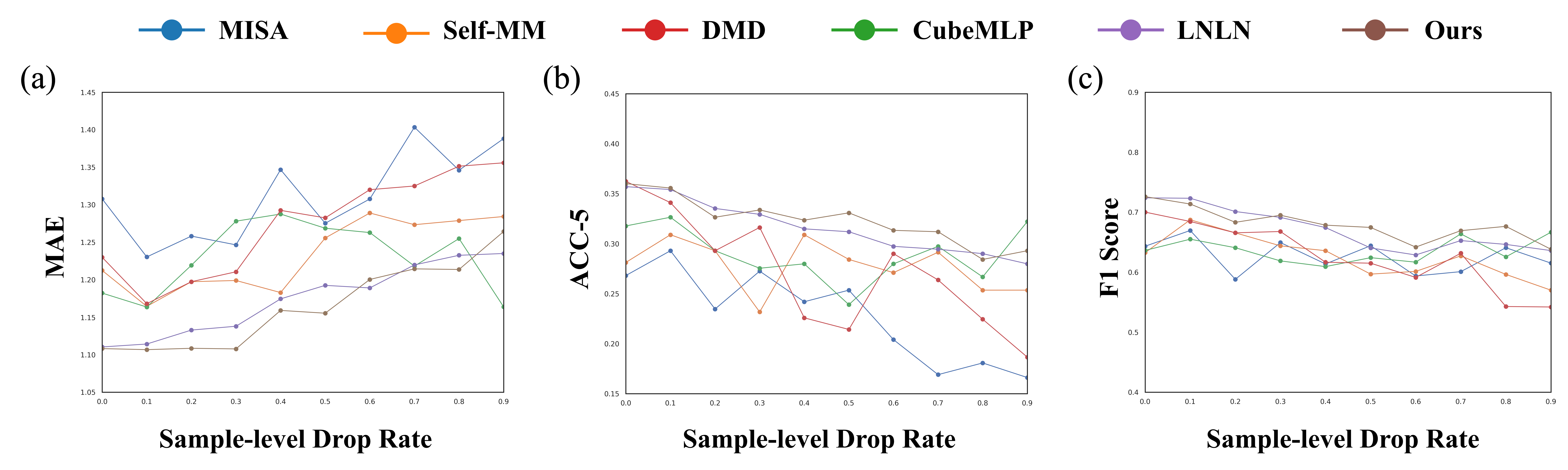}
\caption{Performance curves for a range of sample-level \textit{drop\_rate} values simulating progressively higher levels of \emph{unknown inter-modality} missingness.
Figures~(a)--(c) show the corresponding MAE, ACC-5, and Non0-F1 scores on MOSI, respectively.}
\label{fig:overall_dr}
\end{figure*}

\begin{table*}[ht]
\small
\centering
\caption{Ablation results of simultaneous inter- and intra-modality missingness on MOSI and MOSEI($drop\_rate=0.5$).}
\label{tab:Ablation results1_mosi}
\begin{tabular}{c c c c c c c c c c c c}
\hline
\multirow{2}{*}{\textbf{Model}}
& \multicolumn{5}{c}{\textbf{MOSI}} 
& \multicolumn{5}{c}{\textbf{MOSEI}} \\
\cline{2-6} 
\cline{8-12}
& MAE($\downarrow$) & Acc-7($\uparrow$) & Acc-5($\uparrow$) & Acc-2($\uparrow$) & F1($\uparrow$) &  
& MAE($\downarrow$) & Acc-7($\uparrow$) & Acc-5($\uparrow$) & Acc-2($\uparrow$) & F1($\uparrow$) \\
\hline
\textbf{Senti-iFusion}   & \textbf{1.1554} & \textbf{0.3105} & \textbf{0.3309} & 0.6738 & \textbf{0.6747} & & \textbf{0.6959} & \textbf{0.4507} & \textbf{0.4559} & 0.7276 &  \textbf{0.7819} \\

w/o IIR  & 1.1740 & \underline{0.3090} & \underline{0.3307} & 0.6692 & \underline{0.6736} & & 0.7370 & 0.4437 & 0.4492 & 0.7177 & 0.7341 \\

w/o IAMF\textsubscript{avg} & 1.1671 & 0.2989 & 0.3251 & \underline{0.6794} & 0.6645 & & 0.7407 & 0.4470 & 0.4529 & \underline{0.7286} & 0.7432 \\


w/o \(\mathcal{L}_{ie}\)  & \underline{1.1576} & 0.2799 & 0.3105 & 0.6616 & 0.6618 & & 0.7225 & 0.4460 & 0.4497 & \textbf{0.7325} & \underline{0.7474} \\

w/o \(\mathcal{L}_{rec}^{enc}\)  & 1.1648 & 0.2959 & 0.3207 & \textbf{0.6944} & 0.7215 & & 0.7334 & 0.4452 & 0.4503 & 0.7144 & 0.7473 \\

w/o \(\mathcal{L}_{rec}^{dec}\)  & 1.1635 & 0.3032 & 0.3280 & 0.6631 & 0.6673 & & 0.7338 & 0.4460 & 0.4505 & 0.7171 & 0.7209 \\

w/o strategy & 1.1613 & 0.2988 & 0.3192 & 0.6540 & 0.6536 & & \underline{0.7224} & \underline{0.4480} & \underline{0.4540} & 0.7243 & 0.7368 \\
\hline
\end{tabular}
\end{table*}

\subsection{Performance and Robustness Analysis}

This section presents a comprehensive robustness analysis of our proposed method on the \textbf{MOSI} and \textbf{MOSEI} dataset under varying sample-level inter-modality drop rates, with additional intra-modality masking applied using a fixed random seed. Results are shown from Table \ref{tab:metric_comparison_drop0.5} to \ref{tab:mode_all} and Figure \ref{fig:overall_dr}, which cover sample-level inter-modality missingness scenarios spanning the entire spectrum from 0.0 (complete) to 1.0 (universal absence of one or more modalities across all samples), evaluated at a granularity of 0.1.

Table \ref{tab:metric_comparison_drop0.5} summarizes the results when \textit{50\%} of the test utterances randomly lack one or two entire modalities, while every utterance also experiences additional intra-modality masking.
On \textsc{MOSI}, \textbf{Senti-iFusion} achieves the lowest regression error and the highest classification scores, surpassing the previous best model (LNLN) by roughly 3–4\% across all discrete metrics and reducing MAE by 3\%.
On \textsc{MOSEI}, our model again leads the classification metrics; although CubeMLP attains a slightly lower MAE, \textbf{Senti-iFusion} offers the best overall trade-off between intensity prediction and categorical accuracy.

From Figure \ref{fig:overall_dr}, we observe that under high missing rates, \textbf{Self-MM} exhibits a pronounced mode collapse: at a missing rate of 0.9, it predicts most samples as \emph{weakly negative}, resulting in deceptively high accuracy. However, this behavior indicates poor discriminative capacity, as the model fails to produce fine-grained predictions and instead converges to a dominant sentiment class regardless of input variation.

In comparison, \textbf{MISA} and \textbf{DMD} display slightly more robust behavior. Their predictions remain largely concentrated in the \emph{negative} to \emph{weak negative} range, showing limited variability but retaining some sensitivity to input differences. On the other hand, models such as \textbf{CubeMLP} and \textbf{LNLN} exhibit stronger class bias, especially under severe missingness, indicating weaker generalization under modality degradation.

\textbf{Senti-iFusion} maintains significantly stronger resilience. Even under a 0.9 missing rate, it continues to produce differentiated sentiment predictions, avoiding the collapse into a single dominant class. This highlights its superior ability to preserve semantic fidelity and maintain prediction diversity despite extreme modality loss, validating its robustness in real-world incomplete scenarios.

A more granular examination of model robustness is presented in Table \ref{tab:mode_all}, which details the performance on the \textsc{MOSI} dataset across all six possible modality-missing patterns (Modes 0–5) under a complete inter-modality corruption scenario ($drop_rate = 1.0$). The results reveal a clear distinction in how models handle extreme missingness. 
Our proposed \textbf{Senti-iFusion} demonstrates superior and remarkably consistent performance, achieving top results in four of the six scenarios. Its advantages are particularly pronounced in challenging conditions where critical non-linguistic cues are absent—specifically when the audio ({a}), visual ({v}), or both non-textual modalities ({l,v}, {a,v}) are missing. This robust performance across diverse absence patterns underscores the effectiveness of our integrity-weighted completion and adaptive fusion mechanism, which successfully mitigates the inherent bias toward any single modality.

In stark contrast, the performance of \textbf{LNLN} exhibits a strong dependence on the availability of textual data. It only manages to outperform other baselines in scenarios where the language modality is fully retained. This pattern underscores a critical limitation of its text-dominated architecture, which fails to adequately compensate for the loss of other modalities through cross-modal inference. Consequently, while \textbf{LNLN} excels in text-rich environments, \textbf{Senti-iFusion} maintains a more balanced and resilient performance profile across the entire spectrum of missingness, proving its generalizability to real-world situations where any combination of modalities can be unavailable. 

\noindent\textbf{Above all}, our Senti-iFusion sets new state-of-the-art results across datasets, metrics, missing data ratios, and modality-drop scenarios, confirming its robustness to both inter- and intra-modality missingness.

\begin{figure*}[htbp]
  \centering
  \subfloat[Epoch 1]{%
    \includegraphics[width=0.23\linewidth]{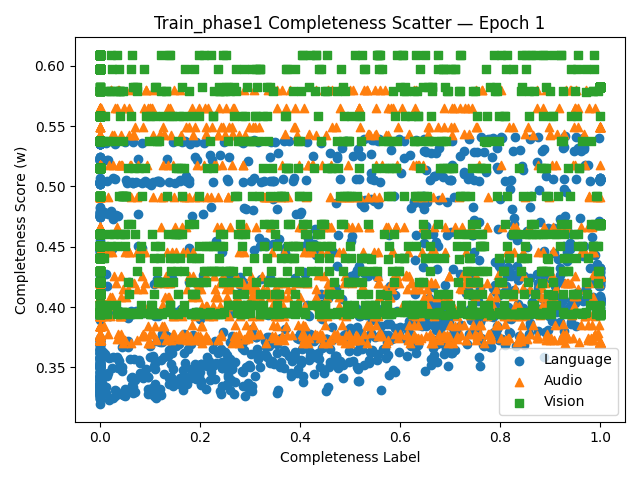}%
  }
  \hfill
  \subfloat[Epoch 15]{%
    \includegraphics[width=0.23\linewidth]{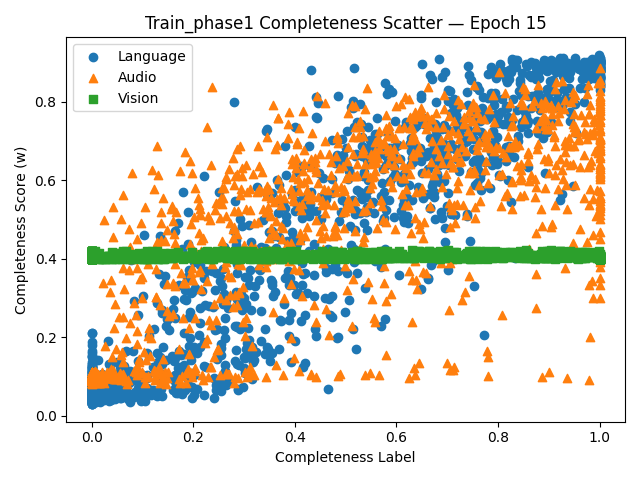}%
  }
  \hfill
  \subfloat[Epoch 35]{%
    \includegraphics[width=0.23\linewidth]{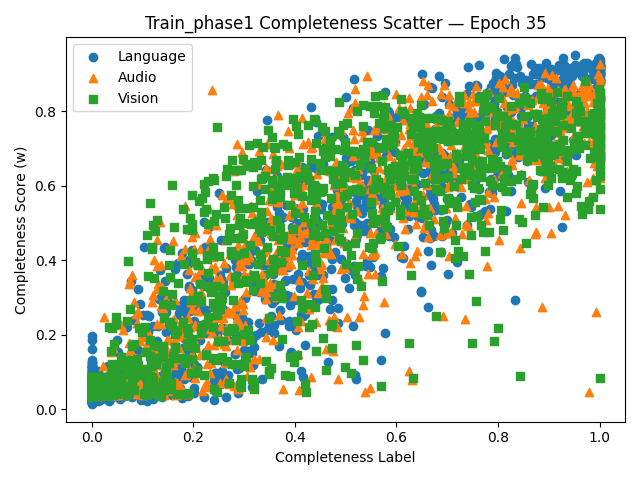}%
  }
  \hfill
  \subfloat[Epoch 60]{%
    \includegraphics[width=0.23\linewidth]{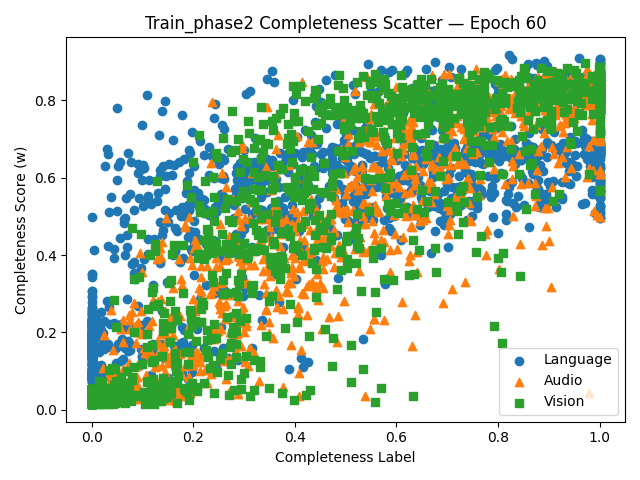}%
  }

  \vspace{0.5em} 
  \subfloat[Epoch 75]{%
    \includegraphics[width=0.23\linewidth]{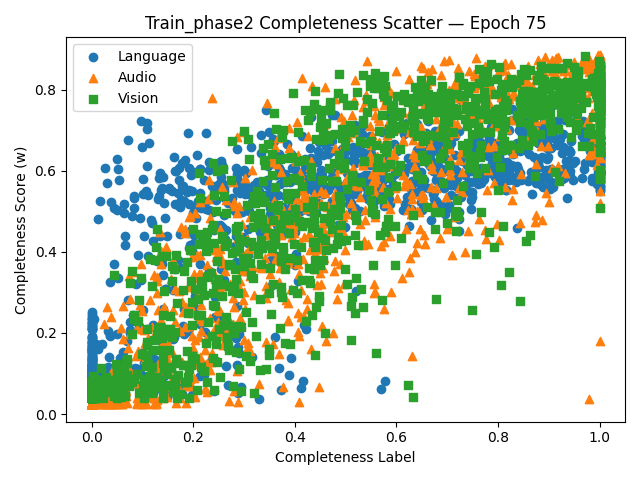}%
  }
  \hfill
  \subfloat[Epoch 100]{%
    \includegraphics[width=0.23\linewidth]{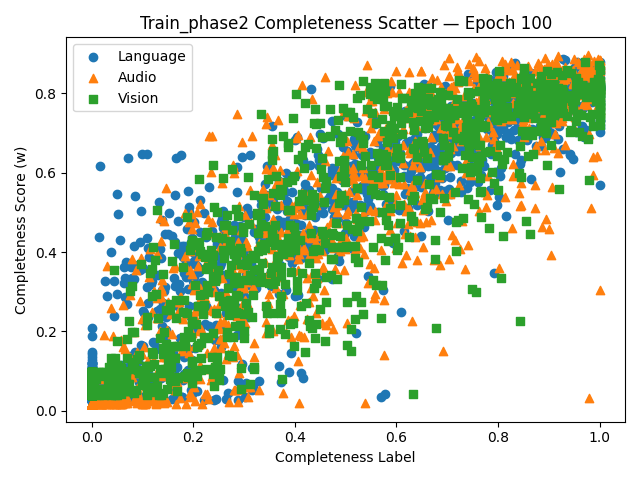}%
  }
  \hfill
  \subfloat[Epoch 120]{%
    \includegraphics[width=0.23\linewidth]{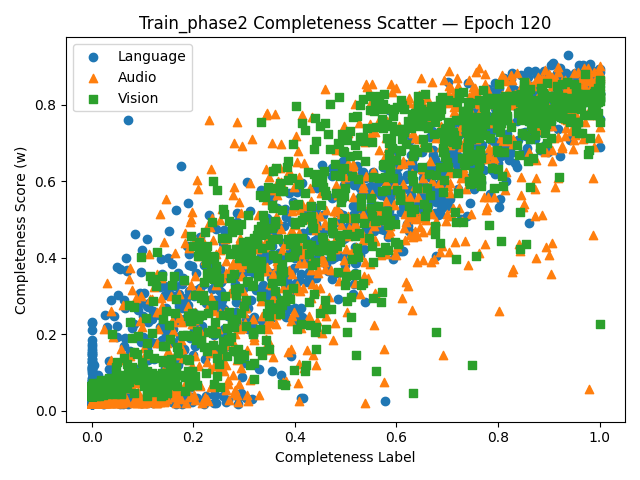}%
  }
  \hfill
  \subfloat[Epoch 150]{%
    \includegraphics[width=0.23\linewidth]{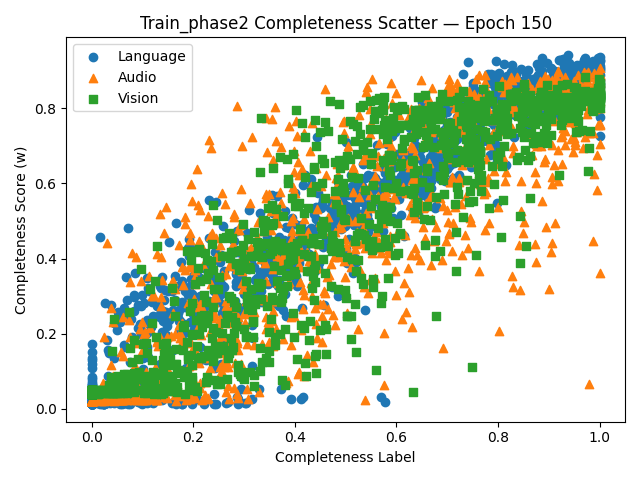}%
  }

  \caption{\textbf{Integrity Estimation Performance during Training.}
  This figure illustrates the prediction behavior of the Integrity Estimation module during training. Each point corresponds to a training-set sample, with the x-axis indicating the ground-truth completeness label and the y-axis showing the predicted score.
  }
  \label{fig:P_completeness_visual_train_logW}
\end{figure*}

\subsection{Ablation Study}

We conducted several ablation experiments to assess the impacts of each component in \textbf{Senti-iFusion} under a 50\,\% inter modality drop rate and random intra-modality masking with fixed seed. The results are summarized in Table\ref{tab:Ablation results1_mosi}.

\subsubsection{Effects of Different Components}

Ablation studies were conducted on \textsc{MOSI} and \textsc{MOSEI} to evaluate the contribution of each module. Removing the integrity estimation loss ($\mathcal{L}_{ie}$) leads to a significant drop in performance, highlighting the importance of integrity-aware modeling. Disabling integrity-weighted completion (w/o IIR) also causes notable degradation (e.g., MAE rises from 1.1554 to 1.1740 on \textsc{MOSI}, and from 0.6959 to 0.7370 on \textsc{MOSEI}), confirming the effectiveness of integrity-based feature retrieval.

Furthermore, removing the dual-level completion loss ($\mathcal{L}_{rec}^{dec}$) leads to in a marked decline in F1, especially on \textsc{MOSEI}, indicating its key role in semantic alignment. Eliminating only $\mathcal{L}_{rec}^{enc}$ results in a smaller performance drop, suggesting that disentangled surrogate generation is most effective when paired with semantic-level supervision. Minor fluctuations in ACC-2 are likely due to random masking or suboptimal hyperparameter configurations across different metrics.

For Feature Disentanglement procedure in the module of \textbf{Integrity-weighted Cross-modal Completion}, we applied two parallel Transformer encoders $\mathcal{E}_m^s$ and $\mathcal{E}_m^p$ to extract disentangled representations. Except for using MAE loss as constraint for the similarity loss $\mathcal{L}_{sim}$, we also conducted experiments using MI loss for similarity encoders and applying a three-way feature constraint to enforce consistent semantic structure features, rather than constraining directly between modality pairs in $\mathcal{M}$, results are presented in Table \ref{tab:comparison_simEncoder_drop0.5}.

Figure~\ref{fig:P_completeness_visual_train_logW} illustrates the predicted versus ground-truth integrity scores across different training epochs for the three modalities (language, acoustic, visual). In the early stages (e.g., Epochs 1 and 15), the predicted scores exhibit high variance and limited alignment with the diagonal line, indicating a noisy and unreliable estimation. As training progresses, the scatter points gradually converge toward the diagonal, demonstrating that the module learns to produce predictions increasingly consistent with ground-truth integrity levels. By Epochs 120 and 150, the predictions are tightly distributed around the diagonal across all modalities, reflecting strong convergence and accurate integrity modeling.

\begin{figure}[h]
\centering
\includegraphics[width=0.95\columnwidth]{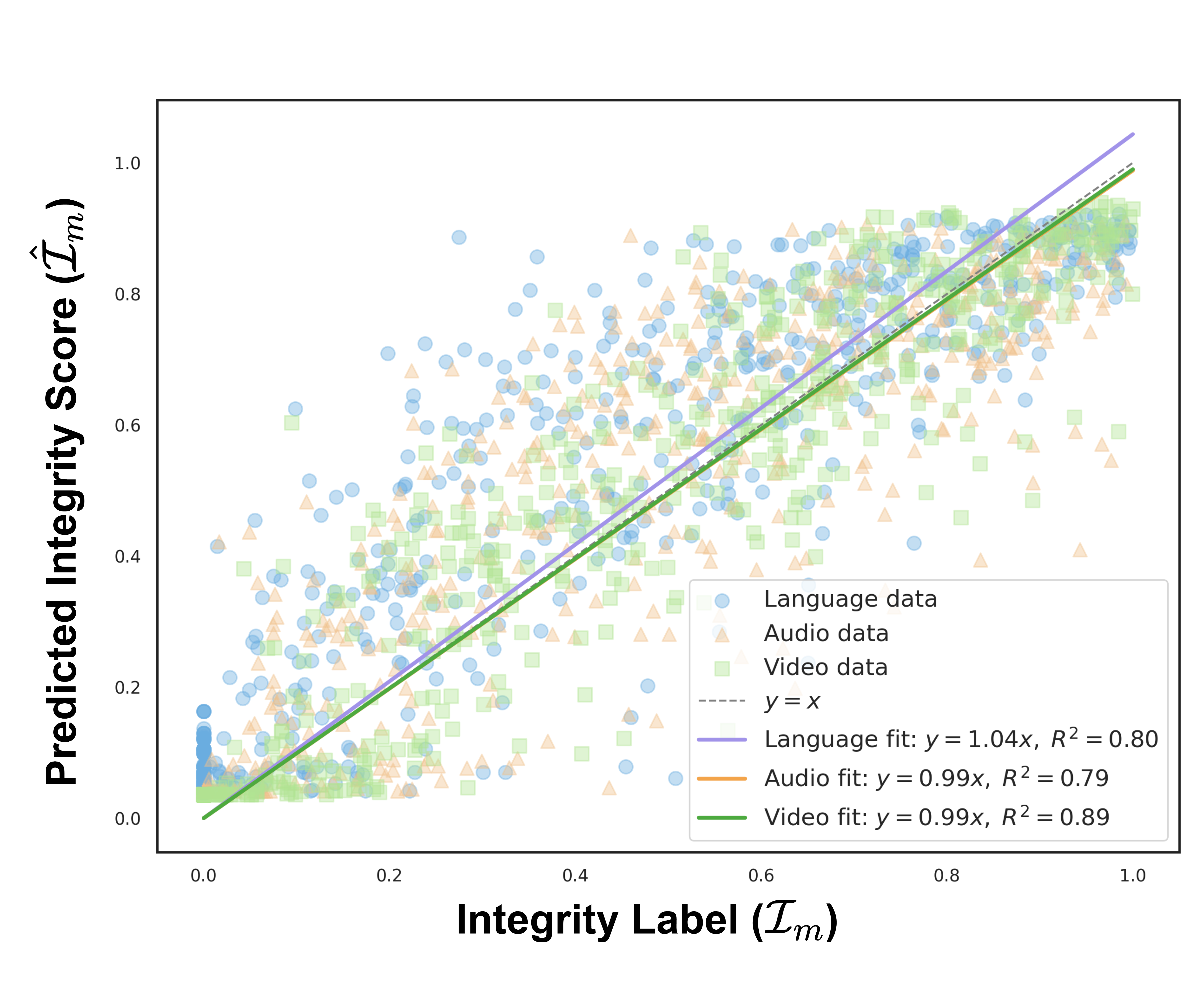} 
\caption{
\textbf{Integrity Estimation Performance during Testing.}
This figure visualizes the prediction performance of the Integrity Estimation (IE) module. Each point represents a test-set sample, where the x-axis denotes the ground-truth integrity label and the y-axis denotes the predicted score produced by the IE module. For each modality—language, acoustic, and visual—we perform linear regression to fit the scatter distribution. The closer the fitted line is to the identity line 
\(y=x\), the more accurate the integrity predictions.
}
\label{fig: completeness_visual_test_logW}
\end{figure}

Figure~\ref{fig: completeness_visual_test_logW} further examines the generalization of the learned estimator on the \textsc{MOSI} test set. For each modality, we plot the predicted integrity scores against the ground-truth values and fit linear regression curves. The resulting coefficients of determination ($R^2$) are 0.80 (language), 0.79 (acoustic), and 0.89 (visual), confirming that the estimator effectively transfers to unseen data. Moreover, the fitted slopes—1.04 (language), 0.99 (acoustic), and 0.99 (visual)—are all close to 1, indicating that the predicted scores not only correlate well with ground-truth but also match their magnitudes, further validating the precision of the integrity estimation module.

\subsubsection{Effects of Training Strategy}
Removing the two-stage progressive training strategy (w/o strategy) consistently harms performance, showing that pretraining integrity estimation stabilizes the subsequent learning process.

\begin{table}[t]
\centering
\footnotesize
\caption{Successful cases where Senti-iFusion significantly improves prediction quality. 
Raw texts are wrapped automatically, and offensive words are omitted.}
\label{tab:case_success_full}

\begin{tabularx}{\columnwidth}{c c X c c}
\toprule
\textbf{ID} & \textbf{True} & \textbf{Raw Text} & \textbf{Ours} & \textbf{Base} \\
\midrule

d3\_k5Xpfmik\_2  & 0.00 & Now Columbo's is basically, um, OCD. & 0.1102 & --1.0060 \\

jUzDDGyPkXU\_3   & 0.00 & It's very reminiscent of District ... & --0.1125 & --1.0143 \\

tmZoasNr4rU\_18  & 0.00 & FFFFLAT [expletive omitted] lip smacking. & --0.1643 & --1.0149 \\

vyB00TXsimI\_1   & 0.00 & But I do wanna tell you about Priest. & --0.1084 & --1.0195 \\

zhpQhgha\_KU\_24 & 0.00 & And not being really stressed out. & --0.1762 & --1.0123 \\

\bottomrule
\end{tabularx}

\end{table}

\subsubsection{Efficiency Analysis}

Compared with the LNLN baseline, Senti-iFusion introduces only a marginal increase in parameters (116.71M vs. 115.97M, +0.6\%) and a small rise in inference latency (2.06 ms vs. 1.01 ms per sample). This minor computational overhead is expected because Senti-iFusion performs an additional surrogate-completion step during feature refinement. Nevertheless, the overhead remains lightweight and does not compromise real-time applicability. Moreover, the accuracy improvements achieved by Senti-iFusion justify the extra computation, demonstrating an effective accuracy–efficiency trade-off.

\subsubsection{Details of Model Implementation}

For Feature Disentanglement procedure in the module of \textbf{Integrity-weighted Cross-modal Completion}, we applied two parallel Transformer encoders $\mathcal{E}_m^s$ and $\mathcal{E}_m^p$ to extract disentangled representations. Except for using MAE loss as constrain for the similarity loss $\mathcal{L}_{sim}$, we also conducted experiments using MI loss for similarity encoders and applying a three-way feature constraint to enforce consistent semantic structure features, rather than constraining directly between modality pairs in $\mathcal{M}$, results are presented in Table \ref{tab:comparison_simEncoder_drop0.5}.

\begin{table}[ht]
\small
\setlength{\tabcolsep}{0.9mm}
\centering
\caption{Comparison of different constraints for $\mathcal{E}_m^{sim}$ on MOSI ($drop\_rate=0.5$).}
\label{tab:comparison_simEncoder_drop0.5}
\begin{tabular}{c c c c c c}
\hline
\textbf{Model} & MAE($\downarrow$) & Acc\_7($\uparrow$) & Acc\_5($\uparrow$) & Acc\_2($\uparrow$) & F1($\uparrow$) \\
\hline
sim\_MI\_loss & 1.1652 & 0.2901 & 0.3163 & 0.6677 & 0.6686\\
sim\_3Way\_feat & 1.1570 & 0.3032 & 0.3280 & \textbf{0.6768} & \textbf{0.6753}\\
\textbf{Senti-iFusion}   & \textbf{1.1554} & \textbf{0.3105} & \textbf{0.3309} & 0.6738 & 0.6747\\
\hline
\end{tabular}
\end{table}

\subsubsection{Case Study}
To qualitatively assess model behavior, we selected eight representative utterances from the \textsc{MOSI} test set where \textbf{Senti-iFusion}'s predictions were notably more accurate than those of a strong baseline. Specifically, we filtered cases where our prediction error was small ($|\text{pred}_{\text{ours}} - \text{true}| < 0.25$) while the baseline error exceeded 1.0. Table~\ref{tab:case_success_full} presents these samples, including raw text and predicted sentiment scores by Epoch 41.

For ID \texttt{tmZoasNr4rU\_\$\_\$18} (“FFFFLAT (offensive word omitted) LIP SMACKING”), annotated as neutral (0.00), the baseline severely overestimated negativity (–1.0149), while our model’s output (–0.1643) was more aligned. A similar observation holds for \texttt{zhpQhgha\_KU\_\$\_\$24} (“AND NOT BEING REALLY STRESSED OUT”), where our model predicted –0.1762 compared to the baseline’s –1.0123.

These examples underscore Senti-iFusion’s enhanced robustness and finer-grained semantic sensitivity—particularly in scenarios with ambiguous or subtle expressions—where conventional models often overreact or misclassify sentiment polarity.

\section{Conclusion}

In this work, we address multimodal sentiment analysis under uncertain modality missingness, where both inter- and intra-modality data absence occur simultaneously. 
Existing methods either tackle only one type of missingness or rely on fixed assumptions, limiting their robustness in real-world applications.
We propose \textbf{Senti-iFusion}, an integrity-centered hierarchical fusion framework that (i) estimates fine-grained completeness scores, (ii) performs integrity-weighted cross-modal completion, and (iii) conducts adaptive fusion by dynamically selecting the most reliable modality.
The dual-depth validation scheme preserves both semantic and statistical fidelity, while the progressive two-stage training strategy ensures stable convergence. 
Experiments on MOSI and MOSEI show that \textbf{Senti-iFusion} outperforms existing baselines and remains robust even under severe missing ratios. 
Ablation studies further confirm the necessity of the key components of the framework.
As a complementary direction, we also aim to investigate how integrity-aware modules like Senti-iFusion can be plugged into large pretrained multimodal backbones to further enhance robustness under missing modalities.

\bibliographystyle{IEEEtran}
\bibliography{references}
\vfill

\end{document}